\documentclass[prl,aps,twocolumn,showpacs]{revtex4}

\usepackage{amsmath}
\usepackage{bm}
\usepackage{graphicx}

\begin{document}

\title{Dynamically Stabilized Bright Solitons in a Two-Dimensional
Bose-Einstein Condensate}

\author{Hiroki Saito}
\author{Masahito Ueda}
\affiliation{Department of Physics, Tokyo Institute of Technology,
Tokyo 152-8551, Japan}

\date{\today}

\begin{abstract}
We demonstrate that matter-wave bright solitons can be stabilized in 2D
free space by causing the strength of interactions to oscillate rapidly
between repulsive and attractive by using, e.g., Feshbach resonance.
\end{abstract}

\pacs{03.75.Fi, 05.30.Jp, 05.45.Yv, 34.50.-s}

\maketitle

A pendulum with an oscillating pivot can have a stable configuration with
its bob situated above the pivot (the inverted pendulum~\cite{Landau}).
This well-known but surprising phenomenon is attributed to a net
stabilizing force produced by alternating stabilizing and destabilizing
forces at a frequency much faster than the natural oscillation frequency
of the fixed-pivot pendulum.
Similar physics has been employed to focus beams of charged particles in
a synchrotron (alternating-gradient focusing~\cite{Feynman}) and to trap
ions in the Paul trap~\cite{Dehmelt}.
In this Letter, we present a novel application of such a stabilizing
mechanism to producing bright solitons in a two-dimensional (2D)
Bose-Einstein condensate (BEC).
By a `bright' soliton we mean a stable, solitary wave whose density
is greater than the background one.

In 1D, when the interatomic interaction is attractive, bright solitons
can be stabilized even without a trapping potential~\cite{Khay,Strecker}.
In 2D or 3D free space, however, the kinetic pressure and the attractive
force cannot balance, and the condensate always collapses or expands.
This can be understood by the following simple argument.
When the characteristic size of the condensate is $R$, the kinetic and
interaction energies are proportional to $R^{-2}$ and $-R^{-d}$ in $d$
dimensions, and an effective potential for $R$ is the sum of these
energies.
The effective potential, therefore, has a minimum only for $d = 1$.
Here we present a novel method to stabilize solitons in 2D by causing the
interaction to oscillate rapidly using, e.g., Feshbach
resonance~\cite{Inouye,Cornish}.

The system considered here is a BEC confined in a quasi-2D axisymmetric
trap~\cite{Gorlitz}, where the axial confinement energy $\hbar \omega_z$
is much larger than the radial confinement and interaction energies.
We assume that the condensate wave function $\psi$ is always in the ground
state of the harmonic potential with respect to the $z$ direction, and
that the dynamics are effectively 2D in the $x$-$y$ plane; we will justify
these assumptions later.
We let the radial confinement frequency $\omega_\perp(t)$ and the s-wave
scattering length $a(t)$ vary in time.
The system is then described by the Gross-Pitaevskii (GP) equation:
\begin{equation} \label{GP}
i \frac{\partial \psi}{\partial t} = -\frac{1}{2} \nabla^2 \psi +
\frac{\omega_\perp^2(t)}{2} r^2 \psi + g(t) |\psi|^2 \psi,
\end{equation}
where $r^2 \equiv x^2 + y^2$ and $g(t) \equiv (8\pi m \omega_z /
\hbar)^{1/2} N a(t)$ describes the strength of interactions.
In Eq.~(\ref{GP}), length, time, frequency, and $\psi$ are measured in
units of $d_0 \equiv [\hbar / (m \omega_{\perp 0})]^{1/2}$,
$\omega_{\perp 0}^{-1}$, $\omega_{\perp 0}$, and $N^{1/2} / d_0$, where
$\omega_{\perp 0} \equiv \omega_\perp(0)$.

We make the strength of interaction oscillate rapidly at frequency
$\Omega$.
The oscillation of the scattering length with $\Omega \simeq \omega_\perp$
is studied in Ref.~\cite{Abdul} in a different context.
Experimentally, the speed of the change in the strength of interaction
using Feshbach resonance is limited by that of the applied magnetic field
$\sim 1$ G/$\mu$s~\cite{Claussen}, which is sufficient for our purpose.
In order to avoid nonadiabatic disturbances that destabilize a soliton
state, we gradually switch on interaction and simultaneously turn off the
radial confinement potential as follows
\begin{eqnarray} \label{tdparam}
g(t) & = & f(t) (g_0 + g_1 \sin \Omega t), \\
\omega_\perp^2(t) & = & 1 - f(t),
\end{eqnarray}
where $f(t)$ is a ramp function
\begin{equation} \label{ramp}
f(t) = \left\{ \begin{array}{ll} t / T & (0 \leq t \leq T) \\
1 & (t > T). \end{array} \right.
\end{equation}
The interaction and the potential can be changed independently
using an optical trap and magnetic-field-induced Feshbach resonance.

\begin{figure}[tb]
\includegraphics[width=8.4cm]{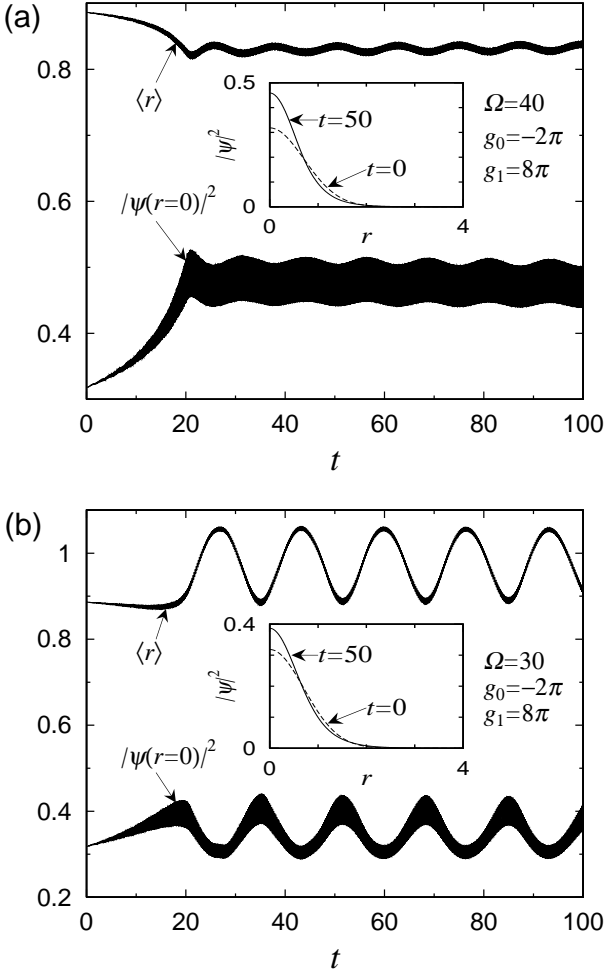}
\caption{
Time evolution of the peak density $|\psi(r = 0)|^2$ and the monopole
moment $\langle r \rangle = \int r |\psi|^2 d{\bf r}$.
The interaction $g(t) = -2\pi + 8 \pi \sin \Omega t$ is switched on and
the radial confinement potential is switched off from $t = 0$ to $20$
according to Eq.~(\protect\ref{ramp}), where $\Omega = 40$ in (a) and
$\Omega = 30$ in (b).
The plots oscillate rapidly at $\Omega$ (which is beyond the resolution
limit of the graph), and the vertical width of the plots represents the
amplitude of the rapid oscillations.
The insets show the density profiles $|\psi(r)|^2$ at $t = 50$ (solid
curves) and $t = 0$ (dashed curves).
}
\label{f:stable}
\end{figure}

We numerically solve the GP equation (\ref{GP}) with time-dependent
parameters (\ref{tdparam})-(\ref{ramp}) using the Crank-Nicholson
scheme~\cite{Ruprecht}.
The initial state is assumed to be the noninteracting ground state in the
presence of the radial confinement potential with $\omega_{\perp 0}$.
We gradually increase the strength of interaction and switch off the trap
according to the linear ramp (\ref{ramp}) with $T = 20$.
Figure~\ref{f:stable} demonstrates the dynamic stabilization of a
soliton in 2D, where the parameters used are $g_0 = -2 \pi$, $g_1 = 8
\pi$, and $\Omega = 40$ [Fig.~\ref{f:stable} (a)] and $\Omega = 30$
[Fig.~\ref{f:stable} (b)].
Since the value of $g_0 = -2\pi$ exceeds the critical strength of
attractive interaction for the collapse $g_{\rm cr} \simeq
-5.8$~\cite{Adhikari}, the condensate would collapse if the oscillating
term $g_1 \sin \Omega t$ were absent.
Even after the radial confinement potential vanishes ($t > 20$), the
soliton shape is maintained as shown in the insets of
Fig.~\ref{f:stable}.
The plots of the peak density $|\psi(r = 0)|^2$ and the monopole moment
$\langle r \rangle = \int r |\psi|^2 d{\bf r}$ rapidly oscillate at
frequency $\Omega$.
We note here that the dynamics of BEC can be separated into two parts: a
rapidly oscillating part with small amplitude and a slow, smoothly varying
part.
After the rapid oscillations are averaged out, the width and the peak
density are fairly constant as seen in Fig.~\ref{f:stable} (a).
However, when the nonadiabaticity of ramp function $f(t)$ is not
negligible, the lowest breathing mode (but no other radial modes) is
excited as illustrated in Fig.~\ref{f:stable} (b), where the breathing
mode is caused by an effective potential due to the oscillating
interaction [see Eqs.~(\ref{Rslow})-(\ref{omegas})].
The amplitude of the excitation decreases with increasing ramp time $T$.
Such excitations can be suppressed by choosing the appropriate ramp
functions for $g(t)$ and $\omega_\perp(t)$.
We have confirmed that no symmetry-breaking mode grows in the azimuthal
direction.
The system is therefore free from such modulational instabilities.

We checked the validity of the quasi-2D approximation by numerically
integrating the GP equation in 3D axisymmetric systems with large
asymmetry parameters $\omega_z / \omega_\perp \sim 50$, and found that
soliton stabilization occurs without axial modes being excited.
Our prediction of dynamically stabilizing bright solitons in 2D is,
therefore, experimentally feasible with an oblate trap as realized by the
MIT group~\cite{Gorlitz}.

To understand the behavior in Fig.~\ref{f:stable}, we employ the
variational method with a Gaussian wave
function~\cite{Garcia,SaitoA,Abdul}
\begin{equation} \label{Gaussian}
\psi(r, t) = \frac{1}{\sqrt{\pi} R(t)} \exp\left[ -\frac{r^2}{2R^2(t)} + i
\frac{\dot R(t)}{2R(t)} r^2 \right],
\end{equation}
where $R(t)$ is the variational parameter characterizing the size of the
condensate, and the imaginary term in the exponent describes mass
current.
Substituting Eq.~(\ref{Gaussian}) into the action that derives the GP
equation (\ref{GP}) and minimizing the action with respect to $R(t)$, we
obtain the equation of motion for $R(t)$ as
\begin{equation} \label{Req}
\ddot{R}(t) = \frac{1}{R^3(t)} + \frac{g_0 + g_1 \sin \Omega t}{2 \pi
R^3(t)},
\end{equation}
where we set $f(t) = 1$.
We separate $R(t)$ into the slowly varying part $R_0(t)$ and the rapidly
oscillating part $\rho(t)$ as $R(t) = R_0(t) + \rho(t)$~\cite{Landau}.
When $\Omega \gg 1$, $\rho(t)$ becomes of the order of $\Omega^{-2}$, and
we keep the terms of the order of up to $\Omega^{-2}$ in the following
analysis.
Substituting this $R(t)$ into Eq.~(\ref{Req}), we obtain the equations of
motion for the rapidly and slowly varying parts as
\begin{eqnarray}
\label{fast}
\ddot{\rho}(t) & = & \frac{g_1}{2\pi R_0^3(t)} \sin \Omega t, \\
\label{slow}
\ddot{R_0}(t) & = & \frac{g_0 + 2\pi}{2\pi R_0^3(t)} - \frac{3 g_1}{2\pi
R_0^4(t)} \overline{\rho(t) \sin \Omega t},
\end{eqnarray}
where the overline indicates the time average of the rapid oscillation.
Equation (\ref{fast}) yields $\rho(t) = -g_1 / [2\pi \Omega^2 R_0^3(t)]
\sin \Omega t$.
Substituting this into Eq.~(\ref{slow}), we obtain the equation of motion
for the slowly varying part as
\begin{equation} \label{Rslow}
\ddot{R_0} = -\frac{\partial}{\partial R_0} \left( \frac{g_0 + 2\pi}{4\pi
R_0^2} + \frac{g_1^2}{16\pi \Omega^2 R_0^6} \right) \equiv -\frac{\partial
U(R_0)}{\partial R_0}.
\end{equation}
The minimum of the effective potential $U$ is attained at
\begin{equation} \label{Rs}
R_{\rm min}^4 = -\frac{3 g_1^2}{4 \pi \Omega^2 (g_0 + 2\pi)},
\end{equation}
and the frequency of small oscillations (breathing mode) around the
minimum is given by
\begin{equation} \label{omegas}
\omega_{\rm br}^2 = \frac{8 \Omega^2}{3 g_1^2} (g_0 + 2\pi)^2.
\end{equation}
The Gaussian approximation thus indicates that a soliton is stable for 
$g_0 < -2\pi$.

The physical mechanism of soliton stabilization can be understood from the
above discussion.
In the inverted pendulum, the interplay between the micromotion of the bob
and the force gradient (i.e., stronger oscillating force for larger
deviation from the equilibrium position) produces a pseudopotential.
Since the pseudopotential is proportional to the square of the amplitude
of the oscillating force~\cite{Landau}, a potential barrier is formed
around the equilibrium position, thereby preventing the pendulum from
swinging down.
Such a mechanism also stabilizes a BEC in a double-well potential with
oscillating interaction~\cite{Ab2000} and an optical beam propagating in
nonlinear medium with an alternating nonlinearity~\cite{Towers}.
In the present case, the oscillating ``force'' for $R$ is given by
$g_1 \sin \Omega t /(2\pi R^3)$, which becomes larger for smaller $R$.
This force gradient and the micromotion of $R$ produce a pseudopotential
proportional to the square of the ``force'' $\propto R^{-6}$, which
prevents the system from collapsing by counteracting the $-R^{-2}$ term
that describes the attractive interaction.

\begin{figure}[tb]
\includegraphics[width=8.4cm]{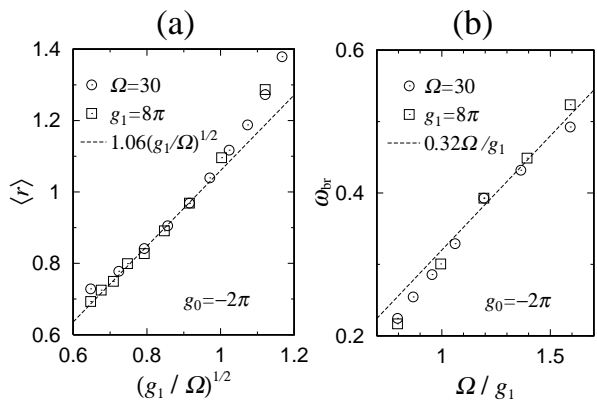}
\caption{
(a) $(g_1 / \Omega)^{1/2}$ dependence of the monopole moment $\langle
r \rangle$ and (b) $\Omega / g_1$ dependence of the breathing-mode
frequency $\omega_{\rm br}$, where $g_0 = -2\pi$.
The circles are obtained by varying $g_1$ for $\Omega = 30$, and the
squares by varying $\Omega$ for $g_1 = 8\pi$.
The dashed lines show $1.06 (g_1 / \Omega)^{1/2}$ in (a) and $0.32 \Omega
/ g_1$ in (b), which are determined to best fit the plots.
}
\label{f:depend}
\end{figure}

In order to stabilize the soliton, $|g_0|$ must exceed the critical value
of collapse, $|g_{\rm cr}| \simeq 5.8$, which is smaller than that of the
Gaussian approximation, $|g_{\rm cr}| = 2\pi$, since the Gaussian wave
function underestimates the peak density~\cite{SaitoA}.
In fact, the GP equation predicts that a soliton state for $g_0 = -2\pi$
is stable (Fig.~\ref{f:stable}).
Although the Gaussian approximation does not accurately describe the exact
soliton stability condition, Eqs.~(\ref{Rs}) and (\ref{omegas}) capture
the $g_1$ and $\Omega$ dependences of $R_{\rm min}$ and $\omega_{\rm
br}$.
Figure~\ref{f:depend} (a) illustrates the monopole moment $\langle r
\rangle$ versus $(g_1 / \Omega)^{1/2}$, where the circles are obtained by
varying $g_1$ and the squares by varying $\Omega$.
We note that the plots are fairly well proportional to $(g_1 /
\Omega)^{1/2}$, in agreement with Eq.~(\ref{Rs}).
Figure~\ref{f:depend} (b) shows the frequencies $\omega_{\rm br}$ of the
slow oscillations [as shown in Fig.~\ref{f:stable} (b)] versus $\Omega /
g_1$.
The plots are linear in $\Omega / g_1$, which is consistent with
Eq.~(\ref{omegas}).

The appropriate parameter range for stabilizing solitons with size $R
\simeq 1$ is found to be $0.4 \lesssim g_1 / \Omega \lesssim 1.2$ and $g_0
\simeq -2\pi$ for $\Omega$ much larger than the breathing-mode frequency
$\omega_{\rm br}$.
For $\Omega \lesssim 10$, the rapid oscillations are disturbed by
nonadiabatic slow dynamics, and the soliton becomes unstable.
In analogy with the inverted pendulum problem, this corresponds to a
situation in which, the pendulum bob falls a long way during a single
pivot cycle at a low pivot oscillation frequency.
Thus, the effective force fails to stabilize the pendulum bob.
When $g_1 / \Omega$ is small or $|g_0|$ is large, the effective force is
not sufficient to prevent the atoms from accumulating at the center.
As a result, the peak density first grows, the condensate then expands
due to the $R^{-6}$ potential, and subsequently most of the expanded atoms
accumulate at the center due to the $-R^{-2}$ potential.
Thus, the condensate repeatedly contracts and expands.
In each expansion, atoms that are elastically scattered with high energy
cannot return to the soliton region due to the absence of the external
confinement potential, and the condensate gradually decays.

\begin{figure}[tb]
\includegraphics[width=8.4cm]{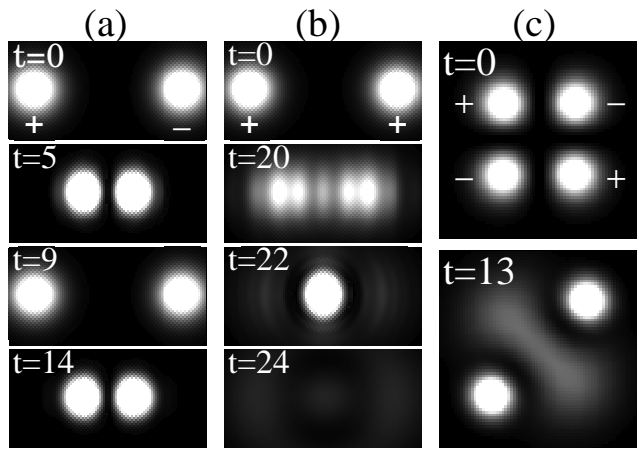}
\caption{
Time evolution of the solitons with $g(t) = -2\pi + 8\pi \sin 40 t$, where
each soliton is prepared as in Fig.~\protect\ref{f:stable} (a).
In (a) and (c), a weak radial harmonic confinement $\omega_\perp^2 = 0.1
\omega_{\perp 0}^2$ is added.
Two solitons are initially placed at $x = \pm 3$ with their phases
differing by $\pi$ in (a) and with the same phase in (b).
In (c), four solitons are initially placed on a slightly deformed square
lattice, such that the adjacent solitons have phases differing by $\pi$.
The images are $8 \times 4$ in (a) and (b), and $8 \times 8$ in (c) in
units of $(\hbar / m \omega_{\perp 0})^{1/2}$.
}
\label{f:interaction}
\end{figure}

We next consider the case in which two or more solitons coexist.
The effective interaction between solitons depends on their relative
phase~\cite{Gordon}.
Figure~\ref{f:interaction} shows the dynamics of solitons with $g(t) =
-2\pi + 8\pi \sin 40 t$.
In Figs.~\ref{f:interaction} (a) and (c), a weak radial confinement
$\omega_\perp^2 = 0.1 \omega_{\perp 0}^2$ is added to keep the solitons
from flying away.
The initial states are taken to be $\psi_{\rm sol}(x + 3, y) \mp \psi_{\rm
sol}(x - 3, y)$ in Figs.~\ref{f:interaction} (a) and (b), where $\psi_{\rm
sol}$ is the wave function of a single soliton prepared as in
Fig.~\ref{f:stable} (a).
When two solitons have phases differing by $\pi$, they repel each other
and oscillate at frequency $\simeq 2 \omega_\perp$
[Fig.~\ref{f:interaction} (a)].
This situation is similar to the quasi-1D experiment performed by the Rice
group~\cite{Strecker}, where adjacent solitons repel each other due to the
$\pi$-phase difference.
When two solitons have the same phase, they approach each other without
the confinement potential, and merge into one condensate, which then
expands [Fig.~\ref{f:interaction} (b)].

The repulsive interaction between solitons in 2D might suggest that
a stable ``soliton lattice'' can be formed without an external periodic
potential.
However, this is not the case.
Figure~\ref{f:interaction} (c) shows time evolution of a ``soliton
lattice,'' where initially the adjacent solitons have phases differing by
$\pi$ and their configuration is slightly perturbed to a rhombus to
examine the stability against lattice distortion.
There is a dynamical instability such that diagonal solitons merge, which
is followed by complicated dynamics.
In the presence of a periodic potential, such alternative-phase structures
are shown to be both stable in 1D~\cite{Bronski} and unstable, but can
survive for a long time with appropriate parameters, in 2D and
3D~\cite{Deconinck}.
In order to study the stability of our system in a periodic
potential, we inserted the ``plug'' (an external potential such as $e^{-4
r^2}$) between diagonal solitons, and found that the lattice structure
survives for a long time.

In conclusion, we have demonstrated that matter-wave bright solitons can
be stabilized in 2D free space by oscillating the strength of the
interaction around an attractive value $g_0 < g_{\rm cr}$ with an
amplitude $g_1 > |g_0|$.
The rapid oscillation of interaction produces an effective barrier that
prevents the condensate from collapsing and stabilizes solitons.
The interaction between solitons can be either repulsive or attractive,
depending on their relative phase, and for repulsive interaction a soliton
lattice can be stabilized using an optical plug.
This novel technique of increasing dimensions of matter-wave bright
solitons might be applied for quantum information processing using BEC
solitons on a microchip substrate~\cite{Hansel}.
It merits further study to examine whether BEC ``droplets'' in 3D can be
created with the appropriate parameters and ramp schemes.

This work was supported by a Grant-in-Aid for Scientific Research (Grant
No. 11216204) and Special Coordination Funds for Promoting Science and
Technology by the Ministry of Education, Science, Sports, and Culture
of Japan, by the Toray Science Foundation, and by the Yamada Science
Foundation.

\end{document}